\RequirePackage[2020-02-02]{latexrelease}
\documentclass[
aps,%
12pt,%
final,%
notitlepage,%
oneside,%
onecolumn,%
nobibnotes,%
nofootinbib,%
superscriptaddress,%
noshowpacs]%
{revtex4}
\usepackage{graphicx}
\usepackage{color}

\begin{document}

\noindent {\it Astronomy Reports, 2023, Vol. , No. }
\bigskip\bigskip  \hrule\smallskip\hrule
\vspace{35mm}


\title{Constraining Supernova Ia Progenitors by their Locations in Host Galactic Disc\footnote{Paper presented at the Fifth Zeldovich meeting, an international conference in honor of Ya.~B.~Zeldovich held in Yerevan, Armenia on June 12--16, 2023. Published by the recommendation of the special editors: R.~Ruffini, N.~Sahakyan and G.~V.~Vereshchagin.}}

\author{\bf \copyright $\:$  2023.
\quad \firstname{A.~A.}~\surname{Hakobyan}}%
\email{artur.hakobyan@yerphi.am}
\affiliation{Center for Cosmology and Astrophysics, Alikhanian National Science Laboratory, 0036 Yerevan, Armenia}%

\author{\bf \firstname{A.~G.}~\surname{Karapetyan}}
\affiliation{Center for Cosmology and Astrophysics, Alikhanian National Science Laboratory, 0036 Yerevan, Armenia}

\author{\bf \firstname{L.~V.}~\surname{Barkhudaryan}}
\affiliation{Center for Cosmology and Astrophysics, Alikhanian National Science Laboratory, 0036 Yerevan, Armenia}

\begin{abstract}
\centerline{\footnotesize Received: ;$\;$
Revised: ;$\;$ Accepted: .}\bigskip\bigskip\bigskip

\scriptsize{Among the diverse progenitor channels leading to Type Ia Supernovae (SNe Ia),
there are explosions originating from white dwarfs with sub-Chandrasekhar masses.
These white dwarfs undergo detonation and explosion triggered by primary
detonation in the helium shell, which has been accreted from a companion star.
The double-detonation model predicts a correlation between the age of the progenitor system and the near peak brightness:
the younger the exploding progenitors, the brighter the SNe.
In this paper, we present our recent achievements on the
study of SNe~Ia properties in different locations within host galactic discs
and the estimation of their progenitor population ages.
Observationally, we confirm the validity of the anticipated correlation
between the SN photometry and the age of their progenitors.}
\end{abstract}

\maketitle

\section{Introduction}
\label{intro}

Type Ia supernova (SN~Ia) is widely believed to originate from a carbon-oxygen white dwarf (WD)
situated in a close binary system.
However, the specific characteristics of the progenitor and the explosion mechanisms leading to SN
are subjects of ongoing debate \cite{2023arXiv230513305L}.
SNe~Ia exhibit a crucial correlation between their $B$-band maximum luminosity and the decline rate of their light curves (LCs; $\Delta m_{15}$):
faster-declining SNe appear fainter \cite{1993ApJ...413L.105P}.
$\Delta m_{15}$ represents the magnitude difference between the SN peak and measurements taken 15 days later.
Considerable efforts have been dedicated to elucidate the origins of SN~Ia progenitors
through investigations of the correlations between SNe~Ia properties and the characteristics of
their host galaxies \cite{2011ApJ...740...92G,2013A&A...560A..66R,2020ApJ...889....8K}.
In particular, the $\Delta m_{15}$ of SN~Ia LCs has been associated with the averaged age
of the host galaxy \cite{2017ApJ...851L..50S}, often used as an approximate indicator of the SN~Ia delay time
(i.e. the time span between progenitor formation and subsequent explosion).
In our recent publication \cite{2020MNRAS.499.1424H}, we demonstrated that the observed correlation
between the $\Delta m_{15}$ of normal SNe~Ia and the global age of their host galaxies seems
to result from the presence of at least two distinct populations.
These populations consist of faster and slower declining SNe~Ia, originating from older and younger stellar populations, respectively.
For the prevailing peculiar SNe~Ia, we showed that subluminous 91bg-like (fast declining) SNe are likely exclusive to
the old population of galaxies, whereas overluminous 91T-like events (slow declining)
solely originate from the young population.
Similar findings have been replicated through more precise age estimations of SNe~Ia host populations,
achieved by employing local properties of SN sites
\cite{2013A&A...560A..66R,2019PASA...36...31P,2019ApJ...874...32R}.

These results can be understood within the context of sub-Chandrasekhar mass
($M_{\rm Ch} \approx 1.4 M_{\odot}$) WD explosion models.
According to this model, the explosion occurs through the double detonation of a sub-$M_{\rm Ch}$ WD,
wherein the detonation of the accreted helium shell triggers a second detonation in
the core of the primary WD \cite{2017ApJ...851L..50S,2010ApJ...714L..52S,2017MNRAS.470..157B}.
SNe~Ia with higher luminosity and slower declining LCs (smaller $\Delta m_{15}$)
are explosions of more massive sub-$M_{\rm Ch}$ WDs.
This link can be attributed to the fact that the SN~Ia luminosity is directly linked to
the mass of $^{56}$Ni synthesized during the explosion \cite{2006A&A...450..241S},
and the amount of $^{56}$Ni, in turn, depends on the mass of the primary WD
(see \cite{2014MNRAS.438.3456P,2018ApJ...854...52S}, for a range of specific models).
Meanwhile, more massive WDs are delivered from more massive main-sequence stars,
which, in turn, possess shorter lifetimes compared to the progenitors of less massive WDs.
Hence, it is reasonable to deduce that the decline rate $\Delta m_{15}$ of SN~Ia
is correlated with the age of the progenitor system giving rise to the SN
\cite{2017ApJ...851L..50S,2021ApJ...909L..18S}.

In the current study, we provide a concise summary of the observational evidences related to the aforementioned scenario,
based on the investigation of the spatial distribution of nearby SNe~Ia within host galactic discs and
estimation of the (dynamical) ages of their progenitor populations through diverse methodologies.
These methods encompass analyzing the correlation between SNe LC decline rates and
the vertical age gradients in galactic discs \cite{2023MNRAS.520L..21B},
assessing distances from host spiral arms \cite{2022MNRAS.517L.132K},
and exploring SN~Ia host stellar population in the star formation desert (SFD) and beyond
\cite{2021MNRAS.505L..52H}.
Our findings offer a valuable new avenue to constrain the nature of SN~Ia progenitors.

\section{SN\lowercase{e} I\lowercase{a} in the star formation deserts of spiral galaxies}

\begin{figure}
\begin{center}$
\begin{array}{@{\hspace{0mm}}c@{\hspace{0mm}}c@{\hspace{0mm}}c@{\hspace{0mm}}c@{\hspace{0mm}}c@{\hspace{0mm}}}
\includegraphics[width=0.18\hsize]{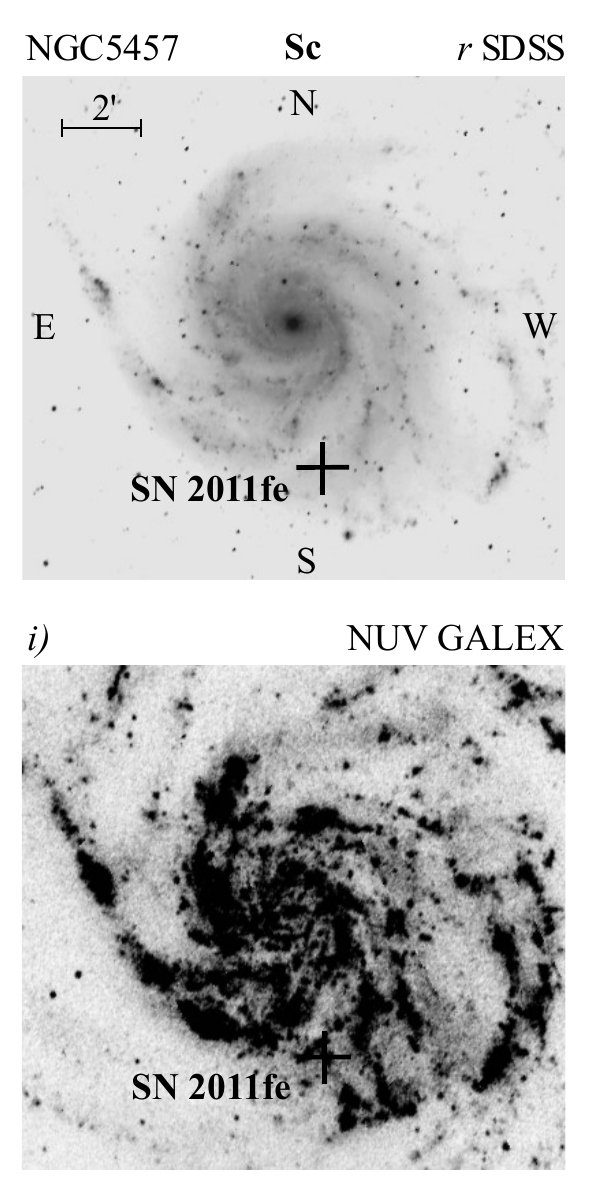} &
\includegraphics[width=0.18\hsize]{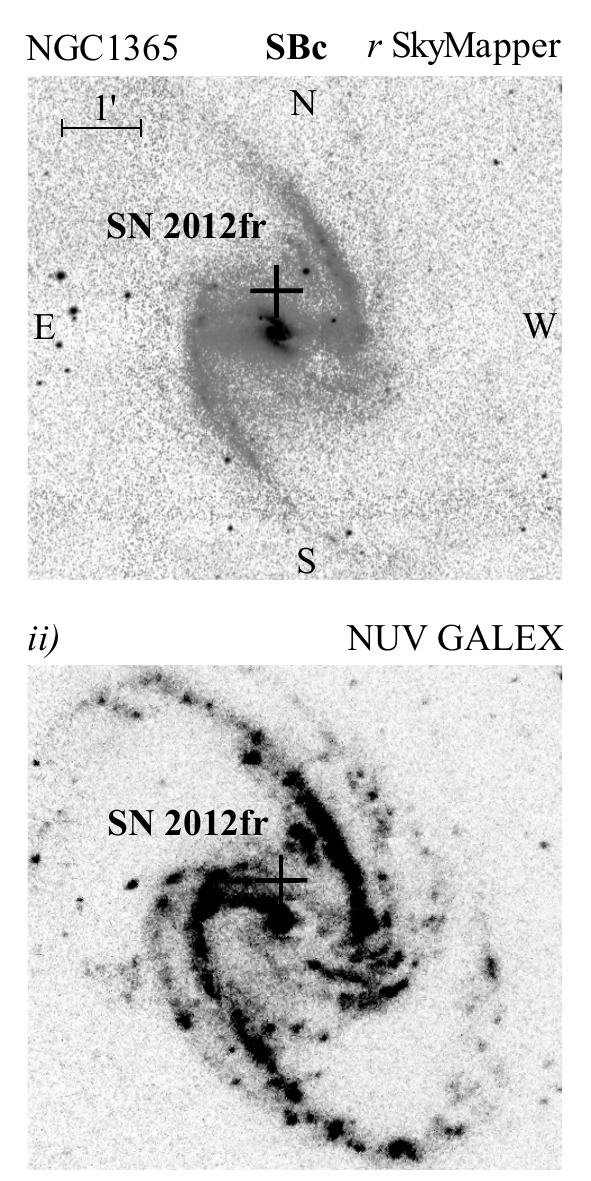} &
\includegraphics[width=0.18\hsize]{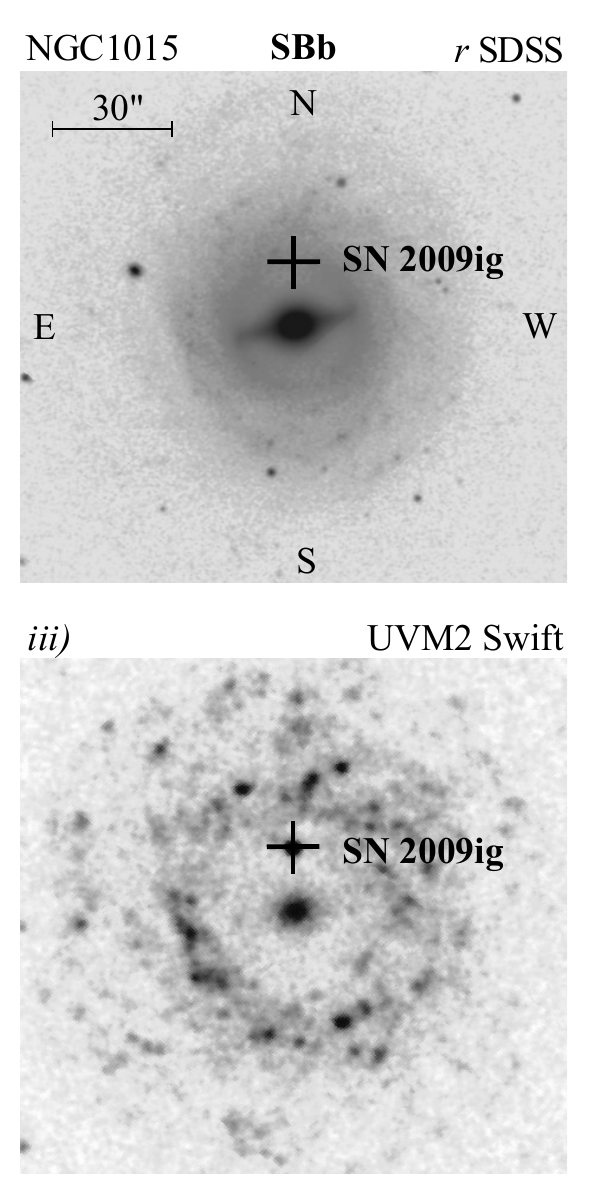} &
\includegraphics[width=0.18\hsize]{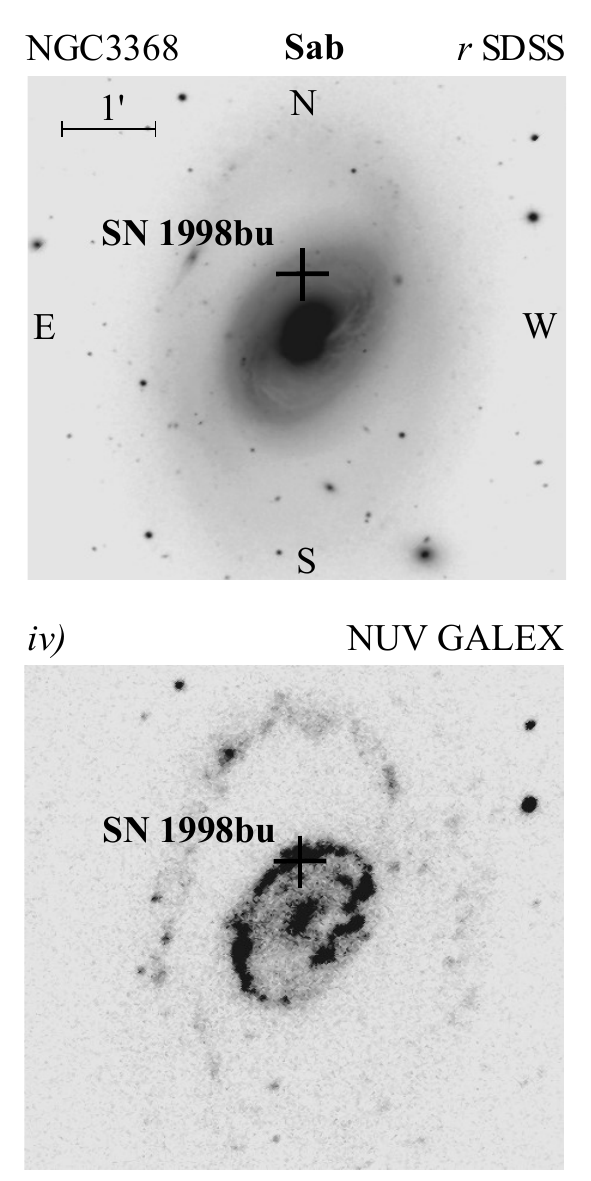}
\end{array}$
\end{center}
\caption{Examples of SN~Ia hosts with different SF discs.
         Top and bottom rows represent optical and UV images, respectively.
         Classes \emph{iii} and \emph{iv} have SFD,
         while classes \emph{i} and \emph{ii} lack an SFD \cite{2021MNRAS.505L..52H}.}
\label{SFSFDexamples}
\end{figure}

The SFD, detected in certain spiral galaxies, refers to a region where a strong bar has
caused sweeping effects, resulting in minimal recent star formation (SF) on both sides of the bar
\cite{2015MNRAS.450.3503J,2018MNRAS.474.3101J}.
Growing evidence from observations and simulations suggests that the SFD primarily comprises old stars.
The suppression of SF in this region is attributed to the formation of the bar
\cite{2019MNRAS.489.4992D,2020A&A...644A..79G},
which dynamically depletes gas from the SFD over a timescale of $\sim2$~Gyr
\cite{2019MNRAS.489.4992D}.
The presence of SF within a bar can manifest in different ways:
SF can occur along the entire length of the bar, exclusively at the ends of the bar,
or the bar itself may exhibit no SF at all.
In certain cases, bars may even dissolve over time,
resulting in the formation of a central SFD within the galactic disc.
Based on the dynamical age constraints of SFDs,
we considered that the delay time distribution (DTD) of its SNe~Ia is
truncated on the younger end, starting from a few Gyr,
in contrast to regions outside the SFD where predominantly young/prompt
SNe~Ia occur (with a delay time of $\sim 500$~Myr,
\cite{2009ApJ...707...74R}).
Given this, and under the assumption that the progenitor's age primarily governs the decline rate,
one would expect the SNe~Ia found in the SFDs should exhibit faster declining LCs.
Below, we effectively demonstrated the soundness of this assumption.

\begin{table}
  \begin{minipage}{100mm}
  \caption{Comparison of the $B$-band $\Delta m_{15}$ distributions between normal SNe~Ia
           in different locations.}
  \tabcolsep 2.1pt
  \scriptsize
  \label{dm15inoutrdem}
    \begin{tabular}{cccccccrr}
    \hline
  \multicolumn{3}{c}{--------- Subsample~1 ---------} & \multicolumn{1}{c}{vs} & \multicolumn{3}{c}{------ Subsample~2 ------} & \multicolumn{1}{c}{$P_{\rm KS}^{\rm MC}$} & \multicolumn{1}{c}{$P_{\rm AD}^{\rm MC}$}\\
  \multicolumn{1}{c}{SN~in} & \multicolumn{1}{c}{$N_{\rm SN}$} & \multicolumn{1}{c}{$\langle \Delta m_{15} \rangle$} && \multicolumn{1}{c}{SN~in} & \multicolumn{1}{c}{$N_{\rm SN}$} & \multicolumn{1}{c}{$\langle \Delta m_{15} \rangle$} && \\
  \hline
     SFD & 12 & 1.32$\pm$0.08 & vs & bar/SF& 12 & 1.07$\pm$0.05 & \textbf{0.005} & \textbf{0.020}\\
     SFD & 12 & 1.32$\pm$0.08 & vs & outer~disc & 52 & 1.13$\pm$0.03 & \textbf{0.009} & \textbf{0.029}\\
     bar/SF & 12 & 1.07$\pm$0.05 & vs & outer~disc & 52 & 1.13$\pm$0.03 & 0.660 & 0.682\\
     SFD$+$bar/SF & 24 & 1.20$\pm$0.05 & vs & outer~disc & 52 & 1.13$\pm$0.03 & 0.445 & 0.477\\
  \hline
  \end{tabular}
  \parbox{\hsize}{\scriptsize{The $P_{\rm KS}^{\rm MC}$ and $P_{\rm AD}^{\rm MC}$ represent the probabilities obtained
                  from the two-sample KS and AD tests, respectively, assessing whether the distributions
                  are derived from the same parent sample.
                  These probabilities were obtained through a Monte Carlo (MC) simulation as explained in \cite{2020MNRAS.499.1424H}.
                  Differences in the distributions that are statistically significant $(P\leq0.05)$ are denoted in bold.}}
  \end{minipage}
\end{table}

\begin{figure}
\begin{center}$
\begin{array}{@{\hspace{0mm}}c@{\hspace{0mm}}}
\includegraphics[width=0.5\hsize]{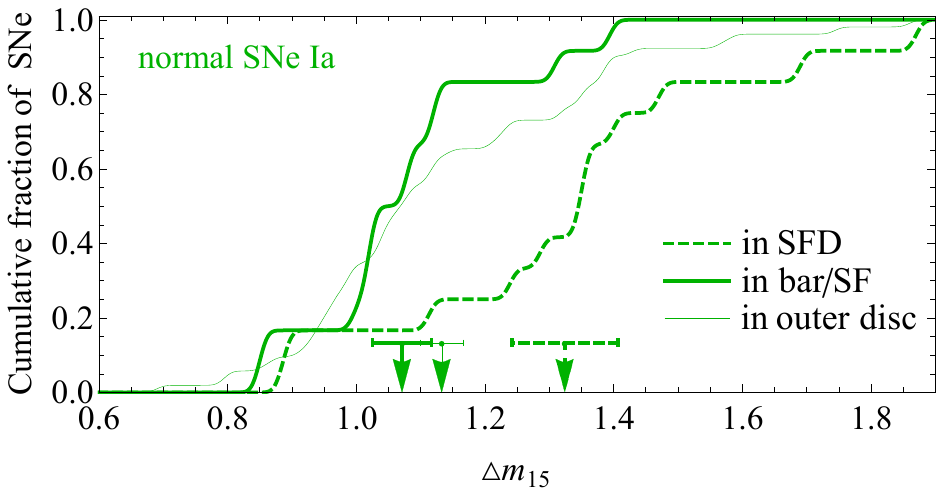}
\end{array}$
\end{center}
\caption{Cumulative $\Delta m_{15}$ distributions for normal SNe~Ia
         in SFD, bar/SF, and in outer disc.
         The mean values, accompanied by their standard errors,
         are depicted by arrows with error bars.}
\label{dm15rsnrdemcum}
\end{figure}

We used a sample of nearby spiral galaxies that hosted a total of 76 normal SNe~Ia
(see \cite{2021MNRAS.505L..52H}, for more details).
We conducted a visual classification of the ionized
(UV and/or H$\alpha$) discs of these galaxies
(see Fig.~\ref{SFSFDexamples}).
Table~\ref{dm15inoutrdem} presents a comparison of the $\Delta m_{15}$ distribution of
normal SNe~Ia within the SFD and those found in the bar/SF (see also Fig.~\ref{dm15rsnrdemcum}).
The results of the Kolmogorov-Smirnov (KS) and Anderson-Darling (AD) tests indicate that
these distributions are significantly different.
Normal SNe~Ia from the SFD, which is predominantly populated by old stars
($\gtrsim$ 2~Gyr; \cite{2019MNRAS.489.4992D}), exhibit, on average, faster declining LCs
in comparison to those found in the bar/SF regions.
In the latter regions, UV/H$\alpha$ fluxes are observed,
suggesting a younger population (age $\lesssim$ a few 100~Myr; \cite{1998ARA&A..36..189K}).

Table~\ref{dm15inoutrdem} further demonstrates that the $\Delta m_{15}$ distribution of
normal SNe~Ia within the outer disc population aligns with the distribution in
the bar/SF but deviates from the distribution observed in the SFD
(see also Fig.~\ref{dm15rsnrdemcum}).
Notably, any inconsistency disappears when we merge the bar/SF and SFD subsamples
and compare the LC decline rates with those observed in the outer disc population
(Table~\ref{dm15inoutrdem}).
These findings indicate that the discs of spiral galaxies predominantly host SNe~Ia
with slower declining LCs ($\Delta m_{15} < 1.25$), which are observed outside
the SFD. The progenitor ages for these SNe~Ia typically peak below 1~Gyr,
aligning with the characteristics of young/prompt SNe~Ia \cite{2014MNRAS.445.1898C}.

Thus, the SFD phenomenon provides an excellent opportunity to distinguish a subpopulation
of SNe~Ia with old progenitors from the broader population residing in host galactic discs,
which comprises both young and old progenitors.
On average, the LCs of this specific SN~Ia subpopulation exhibit faster decline rates,
whose DTD is most likely truncated on the younger side,
beginning from a several Gyr ($\gtrsim$ 2~Gyr).

\section{SN\lowercase{e} I\lowercase{a} distances from spiral arms}

By examining the distribution of SNe~Ia relative to the spiral arms of galaxies,
important links between the host stellar population and the properties of SNe~Ia progenitors can be revealed
\cite{2005AJ....129.1369P,2016MNRAS.459.3130A}.
It is noteworthy that, in accordance with the spiral density wave (DW) theory
\cite{1964ApJ...140..646L,1969ApJ...158..123R},
SF predominantly takes place at shock fronts near the edges of spiral arms.
From these regions, newly born SN progenitors travel in alignment with the disc's rotation direction
relative to the spiral pattern until they reach their respective explosion sites
\cite{2007AstL...33..715M,2016MNRAS.459.3130A}.
The distance from the spiral arm (from the progenitor birthplace) serves
as a potential indicator of SN~Ia progenitor's lifetime,
and as a result, it can be used to constrain the characteristics of SN~Ia progenitors.
In \cite{2022MNRAS.517L.132K} we presented, for the first time, an intriguing approach that offers another valuable means
to investigate the properties of SN~Ia progenitors.

\begin{figure}
\begin{center}$
\begin{array}{@{\hspace{0mm}}c@{\hspace{0mm}}}
\includegraphics[width=0.5\hsize]{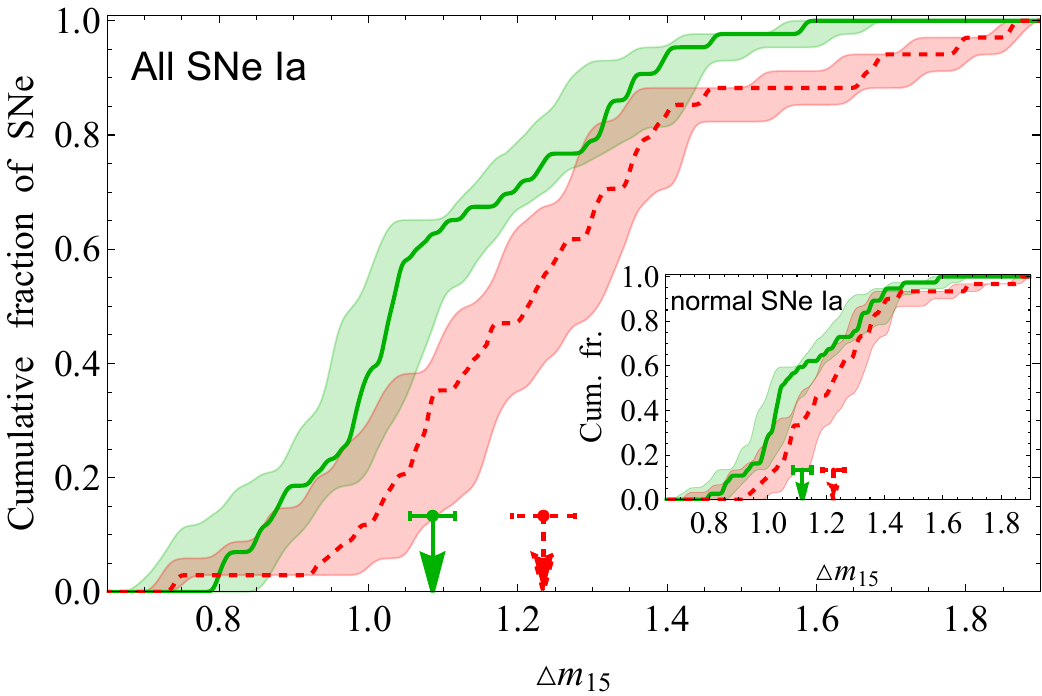}
\end{array}$
\end{center}
\caption{Cumulative $\Delta m_{15}$ distribution of all SNe~Ia observed on spiral arms
         is represented by the green solid curve, while the red dashed curve corresponds to
         SNe located in the interarm regions.
         The corresponding spreads for each curve are depicted by colored regions,
         considering the $\Delta m_{15}$ uncertainties.
         Mean values, along with their standard errors, are indicated by arrows.
         The inset presents a similar representation, but exclusively for normal SNe~Ia.}
\label{Normdm15ArmInterarm}
\end{figure}

For this investigation, we used spiral galaxies hosting 77 SNe~Ia and
our original measurements of the SN distances from the nearby spiral arms
(see \cite{2022MNRAS.517L.132K}, for more details).
Fig.~\ref{Normdm15ArmInterarm} displays the cumulative $\Delta m_{15}$ distributions of
all sampled SNe~Ia in both the \emph{arm} and \emph{interarm} regions.
The $\Delta m_{15}$ values of \emph{arm} SNe~Ia are, on average, smaller (slower declining LCs)
in comparison with those of \emph{interarm} SNe~Ia (faster declining LCs).
The $P$-values obtained from the KS and AD tests presented in Table~\ref{S1S2PKSPAD} strongly suggest
that the two $\Delta m_{15}$ distributions exhibit significant differences.

\begin{table}
  \centering
  \begin{minipage}{93mm}
  \caption{Comparison of the $\Delta m_{15}$ distributions of \emph{arm} and \emph{interarm} SNe~Ia.}
  \tabcolsep 2.2pt
  \scriptsize
  \label{S1S2PKSPAD}
    \begin{tabular}{lccccccc}
    \hline
  \multicolumn{1}{l}{SN} & \multicolumn{1}{c}{$N_{\rm arm \, SN}$} & \multicolumn{1}{c}{$\langle \Delta m_{15} \rangle$} & \multicolumn{1}{c}{vs} & \multicolumn{1}{c}{$N_{\rm interarm \, SN}$} & \multicolumn{1}{c}{$\langle \Delta m_{15} \rangle$} & \multicolumn{1}{c}{$P_{\rm KS}^{\rm MC}$} & \multicolumn{1}{c}{$P_{\rm AD}^{\rm MC}$} \\
  \hline
   All & 43 & 1.09$\pm$0.03 & vs & 34 &  1.23$\pm$0.04 & \textbf{0.006} & \textbf{0.005}\\
   Normal & 37 & 1.12$\pm$0.03 & vs & 30 &  1.21$\pm$0.04 & \textbf{0.037} & 0.075\\
  \hline
  \end{tabular}
  \parbox{\hsize}{The explanations for $P$-values are the same as in Table~\ref{dm15inoutrdem}.}
  \end{minipage}
\end{table}

The aforementioned findings can be interpreted
within the framework of DW theory \cite{1964ApJ...140..646L,1969ApJ...158..123R}
and WD explosion models with a sub-$M_{\rm Ch}$
\cite{2017ApJ...851L..50S,2021ApJ...909L..18S}.
Stars (including SN~Ia progenitors) are believed
to have been born around the shock fronts of spiral arms (birthplace) and subsequently
migrate in the same direction as the disc's rotation, relative to the spiral pattern.
In contrast to SNe~Ia located in the arm regions, those situated in the interarm regions
are expected to have, on average, longer progenitor lifetimes.
This is because they need more time to traverse from their birthplace through the entire
arm before eventually exploding in the interarm regions.
As a result, one can hypothesize that interarm SNe~Ia originate from older progenitors compared
to those found in the arm regions.
Consequently, the arm/interarm separation provides an effective method to differentiate,
on average, between younger and older SN~Ia progenitors.
On the other hand, in sub-$M_{\rm Ch}$ explosion models
\cite{2010ApJ...714L..52S,2017MNRAS.470..157B}
the $\Delta m_{15}$ of SN~Ia is correlated with the age of the progenitor system
(larger $\Delta m_{15}$ values - older progenitors \cite{2017ApJ...851L..50S,2021ApJ...909L..18S}).
Intriguingly, we have observed that the interarm SNe~Ia, on average,
displayed faster declining LCs (larger $\Delta m_{15}$ values), aligning with the expectation mentioned above.

\section{SN\lowercase{e} I\lowercase{a} heights in edge-on galaxies}

\begin{table}
  \centering
  \begin{minipage}{88mm}
  \caption{Comparison of the $|V|/R_{25}$ distributions between different SN~Ia subclasses.}
  \tabcolsep 2.5pt
  \scriptsize
  \label{UR25VR25}
    \begin{tabular}{lcclccc}
    \hline
  Subsample~1 & \multicolumn{1}{c}{$N_{\rm SN}$} & vs & \multicolumn{1}{c}{Subsample~2} & \multicolumn{1}{c}{$N_{\rm SN}$} & \multicolumn{1}{c}{$P_{\rm KS}^{\rm MC}$}&\multicolumn{1}{c}{$P_{\rm AD}^{\rm MC}$}\\
  \hline
    $|V|/R_{25}$ of Normal & 144 & vs & $|V|/R_{25}$ of 91bg & 23 &  0.079 & \textbf{0.010}\\
    $|V|/R_{25}$ of Normal & 144 & vs & $|V|/R_{25}$ of 91T & 30 &  0.685 & 0.588\\
    $|V|/R_{25}$ of 91bg & 23 & vs & $|V|/R_{25}$ of 91T & 30 &  \textbf{0.033} & \textbf{0.022}\\
  \hline
  \end{tabular}
  \scriptsize
  \parbox{\hsize}{The mean heights are $0.07^{+0.01}_{-0.01}$, $0.05^{+0.03}_{-0.02}$,
                  and $0.14^{+0.08}_{-0.04}$ for normal, 91T-, and 91bg-like SNe, respectively.
                  The explanations for $P$-values are the same as in Table~\ref{dm15inoutrdem}.}
  \end{minipage}
\end{table}
\begin{figure}
\begin{center}$
\begin{array}{@{\hspace{0mm}}c@{\hspace{0mm}}}
\includegraphics[width=0.5\hsize]{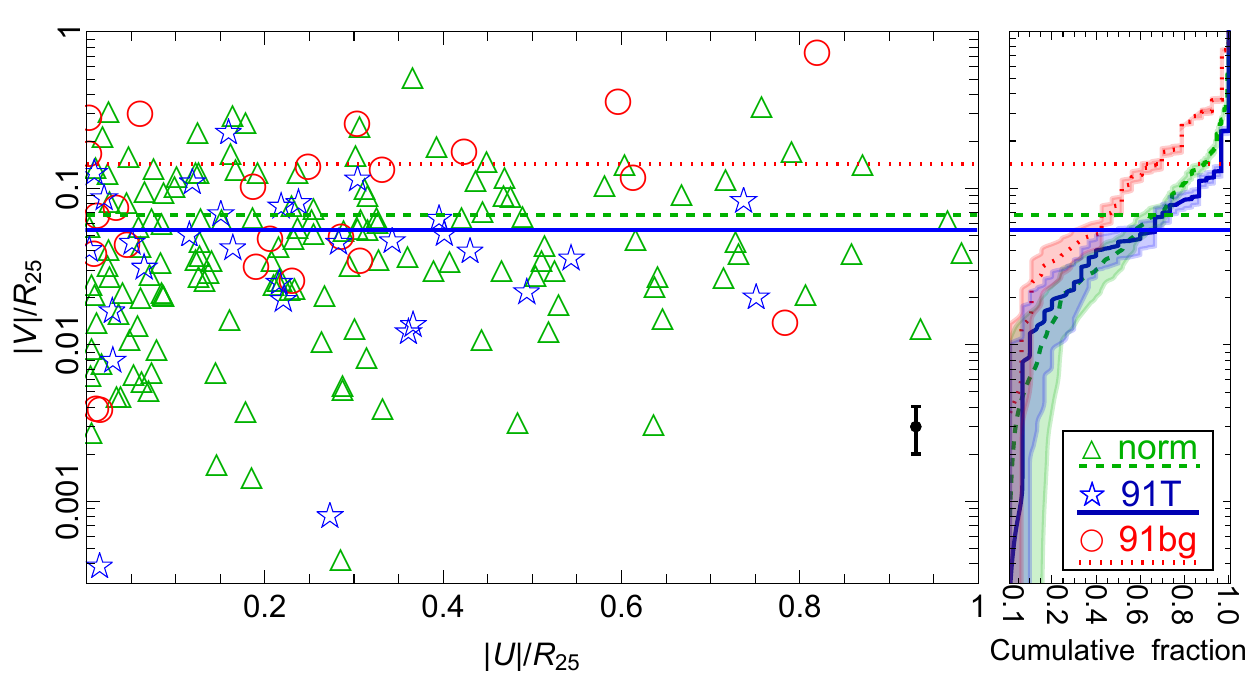}
\end{array}$
\end{center}
\caption{Left panel: vertical $|V|/R_{25}$ vs radial $|U|/R_{25}$ distributions
        for SNe~Ia. The error bar displayed on the right side of the panel
        represents the characteristic error in the height estimation.
        The lines show the mean $|V|/R_{25}$
        values for each SN~Ia subclass. Right panel: the heights' cumulative distributions for
        different SNe~Ia.
        The colored regions surrounding each curve indicate the corresponding spreads,
        taking into account the height uncertainties.}
\label{VR25UR25VR25CDF}
\end{figure}

In \cite{2017MNRAS.471.1390H}, considering the height from the galactic disc
as an indicator of stellar population age
\cite{2005AJ....130.1574S,2006AJ....131..226Y,2018MNRAS.475.1203C},
we demonstrated that the majority of SNe~Ia
are concentrated in the discs of edge-on galaxies, with an approximately two times larger scale height
compared to core-collapse SNe, whose progenitors exhibit ages of up to $\sim$100 Myr.
Furthermore, we demonstrated that the scale height of SNe~Ia is consistent
with that of the older thick disc population observed in the Milky Way galaxy.
Now, we have undertaken an investigation aimed at examining distinct
subclasses of SNe~Ia separately.
This was achieved by analyzing the distributions of nearby 197
normal, 91T-, and 91bg-like SNe heights from the host edge-on discs
(the $V/R_{25}$ heights are normalized to the disc sizes \cite{2023MNRAS.520L..21B}).
Thus, a possible connections between photometric characteristics of SNe~Ia,
such as LC decline rates $(\Delta m_{15})$, and the heights of SNe can be explored.

\begin{figure}
\begin{center}$
\begin{array}{@{\hspace{0mm}}c@{\hspace{0mm}}}
\includegraphics[width=0.5\hsize]{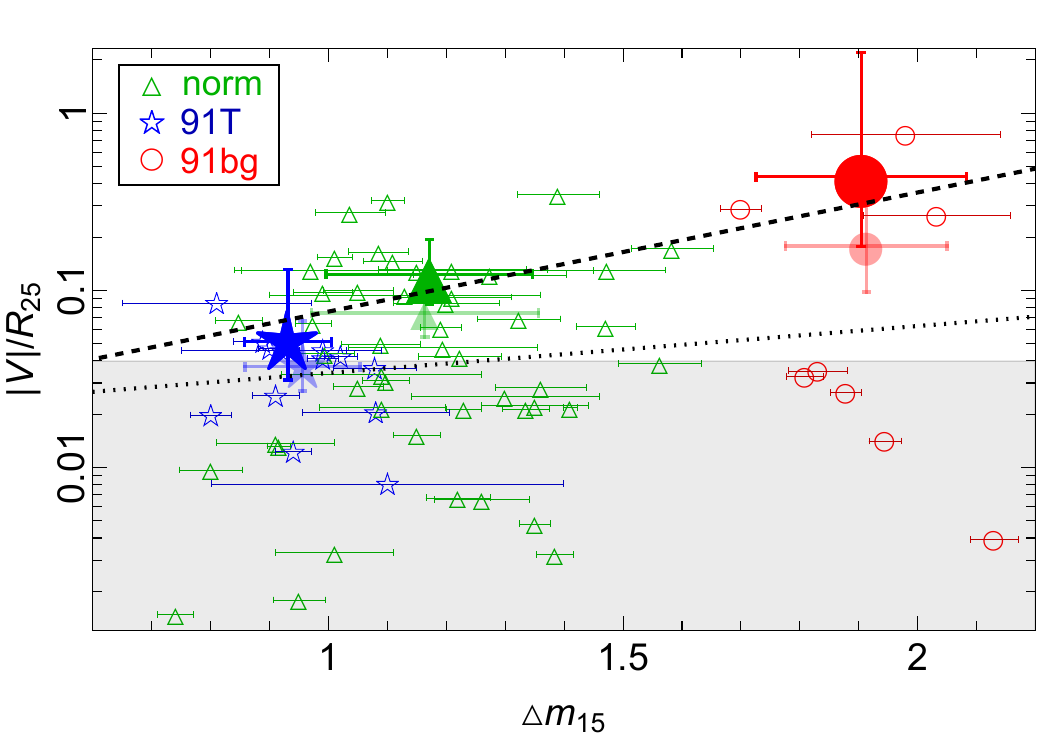}
\end{array}$
\end{center}
\caption{Distributions of $|V|/R_{25}$ vs $\Delta m_{15}$ for the SN~Ia subclasses.
        The dotted and dashed lines, covering all SN~Ia subclasses,
        represent the best-fitting lines for entire and
        dust-truncated discs (outside the shaded area), respectively.
        Averaged values of the parameters (with their errors)
        for entire and dust-truncated samples are depicted by
        medium and large symbols, respectively.}
\label{VR25m15}
\end{figure}

Table~\ref{UR25VR25} statistically illustrates that normal, 91T- and 91bg-like SNe~Ia
exhibit distinct distributions concerning their positions relative to the plane of their host discs.
The mean heights show a gradual increase, beginning with 91T-like events and progressing
through normal and 91bg-like SNe~Ia (see also Fig.~\ref{VR25UR25VR25CDF}).
Recall that spiral galaxies exhibit a vertical stellar age gradient,
where the stellar age tends to increase as the vertical distance from the disc plane increases
\cite{2005AJ....130.1574S, 2006AJ....131..226Y,2018MNRAS.475.1203C}.
Hence, from the perspective of the vertical distribution (as an age tracer),
it can be inferred that the progenitors of 91T-like events are relatively younger
compared to those of normal SNe~Ia, and the progenitors of normal SNe~Ia,
in turn, are younger than those of 91bg-like SNe.
Furthermore, we demonstrated that the SN~Ia decline rates exhibit a correlation
with their heights from the host disc (following the exclusion of selection effects
introduced by dust extinction \cite{2023MNRAS.520L..21B}, see Fig.~\ref{VR25m15}).
The observed correlation aligns with the explosion models involving a sub-$M_{\rm Ch}$ mass WD
\cite{2017MNRAS.470..157B,2017ApJ...851L..50S,2021ApJ...909L..18S}
and the vertical age gradient of the stellar population within discs
\cite{2005AJ....130.1574S, 2006AJ....131..226Y,2018MNRAS.475.1203C}.

\section{Summary}

In \cite{2021MNRAS.505L..52H,2022MNRAS.517L.132K,2023MNRAS.520L..21B},
our focus lied on establishing age constraints for SN~Ia progenitors based on
the dynamical properties of their host galaxies.
We devised novel approaches to assess the DTDs of the progenitors for the subclasses of SN~Ia,
using the vertical age gradients in galactic discs,
distances from host spiral arms,
and locations in the SFD and beyond.
With this in mind, our investigation of SNe~Ia LC decline rates at
different locations within their host galaxies may aid in distinguishing between
SNe~Ia with young progenitors (slow-decliners),
corresponding to the ``prompt'' component with short delay times,
and those (fast-decliners) with ``older'' components exhibiting long delay times.
The observational findings discussed above are in line with the
SN~Ia explosion models involving a sub-$M_{\rm Ch}$ mass WD,
where the SN LC decline rate serves as a suitable indicator of progenitor population age.

Given the limited sizes of our samples,
we strongly encourage to undertake new statistically more powerful studies,
utilizing larger and more robust datasets of SNe~Ia and their hosts.
For better constraints on the nature of SN~Ia progenitors,
we suggest considering integral field observations
using the SFD phenomenon, edge-on galaxies, and spiral hosts with
available details of spiral structure.
Fortunately, ongoing robotic telescope surveys at various locations across the globe,
such as the All Sky Automated Survey for SuperNovae, along with the future
Vera~C. Rubin Observatory (the Large Synoptic Survey Telescope),
will provide the opportunity to realize this goal.

\section*{Funding}

\scriptsize{The work was supported by the Science Committee of RA,
in the frames of the research project \textnumero~21T--1C236.}


\end{document}